\documentclass[12pt,letterpaper,12pt,sort&compress]{extarticle}
\usepackage[T1]{fontenc}
\usepackage[utf8]{inputenc}
\usepackage{array}
\usepackage{booktabs}
\usepackage{amsmath}
\usepackage{graphicx}
\usepackage[numbers,sort&compress]{natbib}

\makeatletter

\providecommand{\tabularnewline}{\\}

\newcommand{\lyxaddress}[1]{
	\par {\raggedright #1
	\vspace{1.4em}
	\noindent\par}
}

\usepackage{indentfirst}
\usepackage{amsfonts}
\usepackage[T1]{fontenc}
\usepackage{ae,aecompl}
\usepackage{sidecap}
\usepackage[section]{placeins}
\usepackage{epsf}

\makeatother

\begin{document}
\title{Self-pinning mechanism for grain boundary stabilization}
\author{Omar Hussein and Yuri Mishin\thanks{Corresponding author. Email: ymishin@gmu.edu}\date{}}
\maketitle

\lyxaddress{Department of Physics and Astronomy, George Mason University, Fairfax,
VA, USA}
\begin{abstract}
Previous research focused on two different mechanisms of microstructure
stabilization in alloys: thermodynamic stabilization by reducing the
grain boundary (GB) free energy and kinetic stabilization by suppressing
the GB mobility by solute drag or embedded pinning particles. Here,
we propose a new GB stabilization mechanism, called self-pinning,
in which the segregation atmosphere of a moving GB spontaneously breaks
into solute-rich clusters, which produce a strong pinning effect in
addition to the free energy reduction resulting from the segregation. The
cluster formation is caused by strong solute-solute attraction at
GBs, leading to a first-order transformation between solute-lean and
solute-rich GB phases. The effect is demonstrated by kinetic Monte
Carlo simulations capturing segregation thermodynamics, GB dynamics,
and solute diffusion. The self-pinning provides an intrinsic stabilization
mechanism for suppressing grain growth that couples thermodynamics
and kinetics. The mechanism obviates the need for pre-existing second
phase inclusions, refocusing the alloy design on GB phase behavior. 
\end{abstract}

\section{Introduction}

Polycrystalline materials are coarsened because the grain boundaries (GBs) possess
excess free energy, which generates a capillary driving force for
grain growth. The grain growth often causes deterioration of the desirable mechanical
and physical properties of the material. This challenge is especially
severe in nanocrystalline alloys, where a large fraction of atoms
reside in GB environments, making thermal stability a central barrier
to maintaining exceptional mechanical and functional properties \citep{Darling:2026aa}.

Two design strategies have been explored to mitigate grain growth,
 termed thermodynamic and kinetic stabilization
mechanisms. Thermodynamic stabilization seeks to reduce the GB free
energy $\gamma_{\mathrm{GB}}$ through solute segregation, thus
lowering the driving force for grain growth through the Gibbs adsorption effect
\citep{gibbs1928collected}. This concept underlies the stability
maps and segregation-based design frameworks for nanocrystalline alloys
\citep{chookajorn2012design,perrin2021stabilized,hussein2024model,HUSSEIN2025121545}.
Kinetic stabilization, in contrast, targets the rate of GB migration.
Classical mechanisms include solute drag, arising from the dissipation
associated with transporting the segregation atmosphere \citep{cahn1962impurity,lucke1972theory,mclean1958grain,sutton1995interfaces,mishin2019solute},
and the Zener pinning by second-phase particles \citep{pa1998five,koju2016zener}.
These mechanisms are often treated as distinct, with thermodynamics
controlling the driving forces and kinetics controlling the GB mobilities.

A key limitation of many classical treatments is the implicit assumption
of \emph{uniform} solute segregation along GBs \citep{cahn1962impurity,lucke1972theory,mclean1958grain,sutton1995interfaces,hillert1976treatment,abdeljawad2017grain},
with only limited consideration of asymmetric segregation profiles
and their influence on drag behavior \citep{alkayyali2021grain}.
In reality, solute segregation is strongly heterogeneous. Special
GBs, GB phases~\citep{MRS-Bulletin-GB-phases, cantwell2014grain,baram2011nanometer,rheinheimer2015non},
facets, and junctions \citep{barnett2024triple} can accumulate large
solute excesses, leading to the formation of nanoscale clusters or
solute-rich interfacial regions \citep{zhou2016grain,kwiatkowski2018phase,darvishi2020segregation,liu2024near,tytko2012microstructural,stoffers2015complex,nikitin2026solid,nikitin2026physical,o2018grain,nenninger2025solute,hu2025stability}.
The clustering fundamentally alters the solid-solution limits \citep{nikitin2026solid},
mechanical strength and deformation mechanisms \citep{liu2024near},
and the stability and mobility of GB disconnections \citep{hu2025stability}.

Analogous to bulk phase transformations, GBs can undergo first-order
transitions between distinct interfacial phases, manifested in discontinuities
in segregation isotherms that reflect abrupt switching between solute-lean
and solute-rich GB states at low temperatures. Analytical and mesoscale
studies \citep{ma2003solute,mishin2019solute,wynblatt2006anisotropy,wynblatt2006some,wynblatt2008interfacial,wynblatt2008solid}
have shown that GB motion can affect the stability and transformations
of low-dimensional phases. GB motion can modify their relative stability
and, in some cases, dynamically stabilize phases that are unstable
under equilibrium conditions \citep{mishin2019solute}. These insights
motivate a unified perspective in which thermodynamic and kinetic
stabilization mechanisms are intrinsically coupled: the same segregation effect
that reduces $\gamma_{\mathrm{GB}}$ can, through interfacial phase
separation, generate solute clusters that strongly resist GB migration.

Here, we propose \emph{self-pinning} as an intrinsic stabilization
mechanism in which GB phase separation produces solute-rich clusters
that act as \emph{in situ} pinning centers, without requiring a pre-existing
particle dispersion. Atomistic simulations have already provided qualitative
evidence that GB-localized heterogeneous solute enrichment can anchor migrating
boundaries and suppress coarsening \citep{koju2018atomistic}. More
recent theoretical developments further suggest that stochastic segregation
and interfacial phase coexistence can sustain persistent heterogeneous
GB states with strong kinetic consequences \citep{mishin2023stochasticI,mishin2023stochasticII}.

Quantitatively capturing the self-pinning requires a modeling framework
that resolves (i) segregation thermodynamics and GB miscibility gaps,
(ii) solute diffusion over long timescales, and (iii) coupled evolution
of diffusion, GB migration, and interfacial phase transformations.
Continuum phase-field models \citep{li2009phase,shahandeh2012friction,gronhagen2007grain,abdeljawad2017grain,kim2008grain}
can describe diffuse interfaces and evolving microstructures, but
they rely on prescribed mobility laws and kinetic parameters. In contrast,
molecular dynamics (MD) simulations \citep{mendelev2002impurity,sun2014direct,rahman2016molecular}
provide atomistic details but are generally limited to timescales
that are too short to represent diffusion of substitutional solute atoms. To bridge
this gap, kinetic Monte Carlo (KMC) frameworks \citep{mendelev2001grain,wicaksono2013three,mendelev2002impurity,mendelev2002domain}
can overcome the intrinsic timescale limitations of MD
while retaining a thermodynamically consistent description of GB segregation
and interfacial phase separation. Our recently developed KMC model
enables systematic exploration of dynamic regimes and self-pinning
conditions that are inaccessible to conventional MD and difficult
to parameterize within continuum models \citep{hussein2024model,HUSSEIN2025121545}.
This KMC approach will be applied in this work to demonstrate the
proposed self-pinning effect.

\section{Methods}

\subsection{The Model}

Full details of the model were reported in our previous  publications
\citep{hussein2024model,HUSSEIN2025121545}. We simulate a two-dimensional
polycrystal on a rectangular square lattice with periodic boundary
conditions. The system contains $N$ sites labeled with an index $k=1,\ldots,N$.
Each site represents a small crystallite with one of $q$ orientations
$\sigma_{k}\in\{1,\ldots,q\}$ in the spirit of a Potts model. The misorientation relative to neighboring
sites is quantified by the number $n_{k}$ of nearest neighbors with
a different orientation, 
\begin{equation}
n_{k}=\sum_{(kl)}\left(1-\delta_{\sigma_{k}\sigma_{l}}\right),\label{eq:2}
\end{equation}
where $\delta_{\sigma_{k}\sigma_{l}}$ is the Kronecker delta and $(kl)$
sums over the four nearest neighbors of the site $k$. GB sites are identified
using the locator function 
\begin{equation}
\phi(n_{k})=1-\frac{1}{4}\left(n_{k}-2\right)^{2},\label{eq:phi}
\end{equation}
which peaks at $\phi=1$ for $n_{k}=2$, which is typical of GB environments.
Crystallographic interactions are modeled by a repulsive nearest-neighbor
penalty $J_{gg}>0$ for unlike orientations, giving 
\begin{equation}
E_{\mathrm{cryst}}=\sum_{k}J_{gg}n_{k},\label{eq:1}
\end{equation}
with $E_{\mathrm{cryst}}=0$ for a perfect single crystal.

Solute atoms are described by a lattice gas with occupation variables
$\xi_{k}\in\{0,1\}$ and a mean concentration $c=\sum_{k}\xi_{k}/N$.
Solute energetics include an inter-grain contribution $J_{s}$ and
an attraction to GBs with strength $J_{sg}<0$: 
\begin{equation}
E_{sg}=\sum_{k}\xi_{k}J_{s}+\sum_{k}\xi_{k}J_{sg}\phi(n_{k}).\label{eq:solute-GB-1}
\end{equation}
Solute--solute interactions are included for nearest-neighbor pairs,
with a bulk interaction parameter $J_{ss}$ and an additional GB-dependent
shift controlled by a parameter $J_{ssg}$: 
\begin{equation}
E_{ss}=\frac{1}{2}\sum_{k}\sum_{(kl)}\xi_{k}\xi_{l}J_{ss}+\frac{1}{2}\sum_{k}\sum_{(kl)}\xi_{k}\xi_{l}J_{ssg}\phi(n_{k,l}),\label{eq:s-s}
\end{equation}
where $n_{k,l}=(n_{k}+n_{l})/2$. This formulation independently tunes
solute interactions inside the grains and within the GB segregation
atmospheres.

To drive microstructural evolution, we create a bias toward a selected
orientation $\sigma_{1}$ by adding another energy term
\begin{equation}
E_{\mathrm{syn}}=\sum_{k}F\delta_{\sigma_{k}\sigma_{1}},\label{eq:synthesis-force}
\end{equation}
where $F$ is a synthetic driving force. Negative (positive) $F$
favors (penalizes) the selected orientation. Thus, the total energy
of a microstate $i=\{\sigma_{k},\xi_{k}\}$ is 
\begin{equation}
E_{i}=E_{\mathrm{cryst}}+E_{sg}+E_{ss}+E_{\mathrm{syn}},\label{eq:E_tot}
\end{equation}
and depends on the model parameters $J_{gg}$, $J_{s}$, $J_{sg}$,
$J_{ss}$, $J_{ssg}$, and $F$.

The time evolution of the system is implemented by a rejection-free ($n$-way)
KMC with harmonic transition state theory rates \citep{vineyard1957frequency}
\begin{equation}
\nu_{ij}=\nu_{0}\exp\!\left(-\frac{\varepsilon_{ij}}{k_{B}T}\right),\label{eq:1.01-1-1}
\end{equation}
using nonlinear barriers \citep{hussein2024model,mishin2023stochasticI,mishin2023stochasticII,HUSSEIN2025121545}
\begin{equation}
\varepsilon_{ij}=\begin{cases}
\varepsilon_{0}\exp\!\left(\dfrac{E_{ij}}{2\varepsilon_{0}}\right), & E_{ij}\le0,\\[6pt]
E_{ij}+\varepsilon_{0}\exp\!\left(-\dfrac{E_{ij}}{2\varepsilon_{0}}\right), & E_{ij}>0,
\end{cases}\label{eq:1.02-1}
\end{equation}
where $E_{ij}=E_{j}-E_{i}$ and $\varepsilon_{0}$ is the unbiased
barrier. Two types of events are implemented: (i) orientation \emph{flips},
where a site changes its orientation to any of the $(q-1)$ alternative
orientations, and (ii) solute \emph{jumps} to an empty nearest-neighbor
site. Flips cause growth or shrinkage of grains, while jumps implement
solute diffusion. Flips use a constant unbiased barrier $\varepsilon_{0}^{g}$.
Solute jumps use a different unbiased barrier $\varepsilon_{0}^{s}$
that is reduced in GBs to capture short-circuit diffusion: 
\begin{equation}
\varepsilon_{0}^{s}=\varepsilon_{00}^{s}\left[1-\eta\phi(n_{k,l})\right],\label{eq:Solute-barrier-1}
\end{equation}
with $0\le\eta<1$ (typically $\eta=1/2$)~\citep{mishin1997grain,chesser2024atomic}.
The actual transition barrier is then biased via Eq.(\ref{eq:1.02-1}).
The solute population is conserved, whereas the site orientations
are unconstrained. Thus, the simulated ensemble is canonical for solute
atoms and grand-canonical with zero chemical potential for the orientations.
At each KMC step, all flip and jump rates are calculated, an event is
selected in proportion to its rate, and the time is advanced by the total
escape rate, following our prior implementation \citep{hussein2024model,HUSSEIN2025121545}.

The simulations are performed in normalized valiables listed in Table~\ref{tab:variables}.
The energies are scaled by $J_{gg}$ and time by $\nu_{0}^{-1}$.
Unless stated otherwise, we take $J_{s}=0$.

\subsection{Qualitative metrics}

\subsubsection{Grain boundary identification and tracking}

GB sites were identified using the structural order parameter $\phi(n)$.
Because a GB has a finite width on the lattice and segregation extends
into the near-boundary neighborhood, we define a \emph{GB region}
by expanding the set of GB sites to include all lattice sites within
two lattice spacings of any GB site (i.e., a $\pm2$-cell neighborhood).
This expanded GB region was used to classify solute atoms into those
located in the GB segregation atmosphere and those located inside
the grains. This classification did not affect the actual simulations.

\begin{figure}[th!]
\centering \includegraphics[width=0.73\columnwidth]{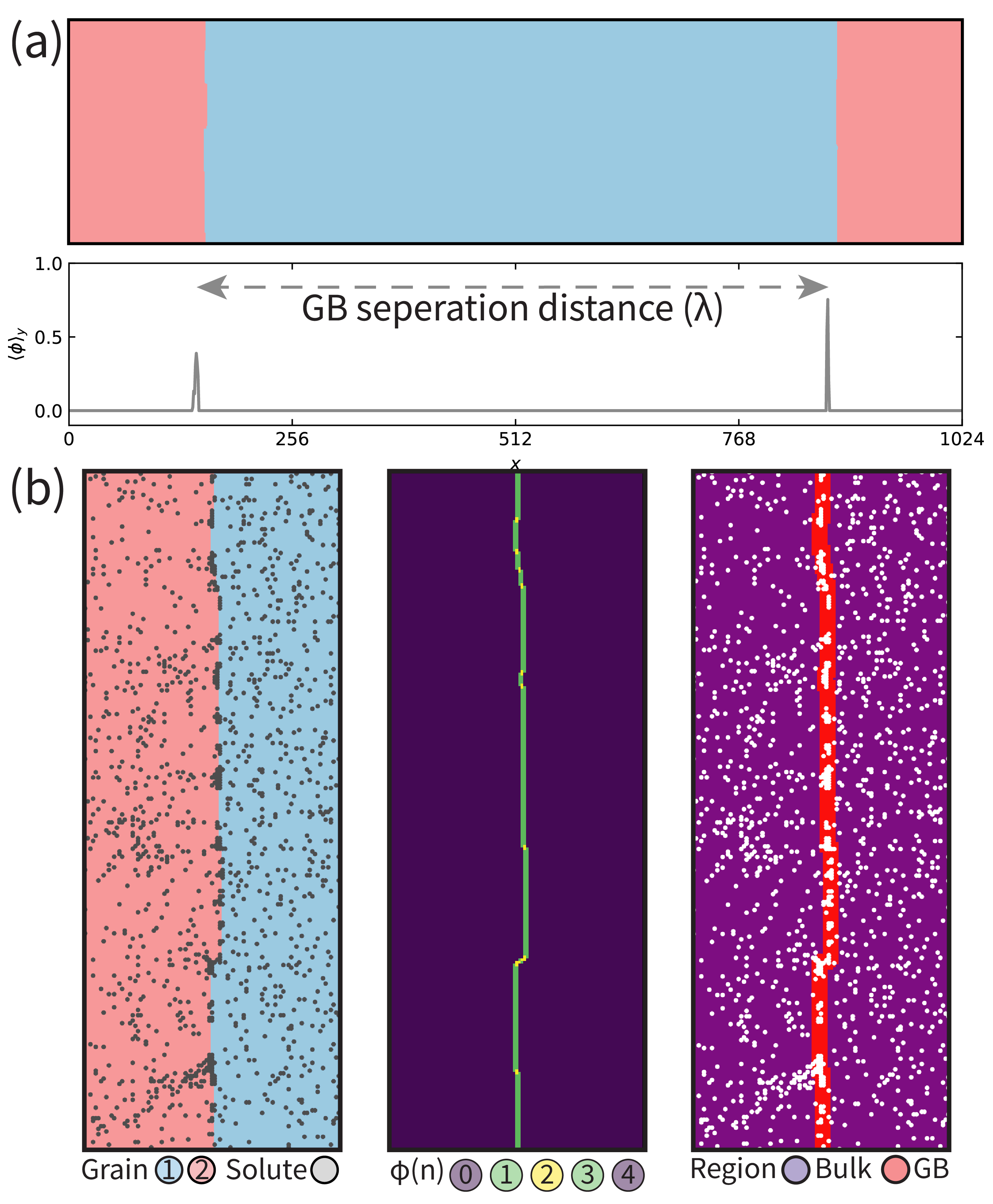} \caption{(a) Illustration of the GB separation metric used in this work. A
snapshot shows the detection of two GBs evolving under an applied
driving force. The GB separation $\lambda$ is defined from the locations
of the peaks of the structural order parameter $\phi(n)$. The lower
panel shows the line-averaged $\langle\phi\rangle_{y}(x)$ profile
with the peaks. (b) Representative configurations illustrating the
correspondence between grain structure, the GB locator $\phi(n)$,
and the solute distribution. Left: grain orientations with solute
atoms overlaid. Center: spatial map of $\phi(n)$ highlighting the
GB core and its local roughness. Right: solute distribution with the
GB region defined as all sites within $\pm2$ lattice spacings of
the GB core, distinguishing the solute atoms in the segregation atmosphere
from those located inside the grains. }
\label{fig:qual_metrics}
\end{figure}

For configurations containing approximately flat GBs, their instantaneous positions were identified from the line-averaged profile $\langle\phi\rangle_{y}(x)$,
where the $x$-axis is normal to the boundaries. The boundaries were
associated with the dominant peaks of the function $\langle\phi\rangle_{y}(x)$,
and their positions were taken as the locations of the peaks. In bicrystals
simulations, the GB separation $\lambda$ was defined as the distance
between the two peaks. This separation provides a scalar
measure of the GB motion over time. Figure \ref{fig:qual_metrics}(a)
illustrates the procedure.

\subsubsection{Grain boundary velocity}

The GB velocity $V$ in bicrystalline samples was determined from
the temporal evolution of the GB separation $\lambda$. To
eliminate transient effects associated with initialization and early
structural relaxation, velocity measurements utilized only the
final portion of each simulation, after the system had reached steady 
state.

To define a time scale relevant to GB migration in the presence of
solute diffusion and GB--solute interactions, we recorded
the accumulated flip time $t_{f}$, defined as the number of accepted
flip events. This clock focuses specifically on the intrinsic GB dynamics.
Importantly, the accumulated flip time captures the effect of the
underlying GB--solute interactions. In cases of strong solute--GB
coupling, a large number of flip events may be accepted while producing
only limited GB displacement. This behavior manifests itself effectively 
as a reduced migration velocity. Conversely, weaker solute--GB interactions
result in fewer accepted flip events for a comparable GB displacement,
corresponding to higher effective velocities. In this framework, the
maximum possible GB velocity corresponds to the idealized limit in
which every accepted flip event advances the GB monotonically in a
certain direction without back-and-forth fluctuations. In this case,
each accepted attempt contributes directly to net GB migration, yielding
the highest attainable velocity for a given flip rate.

GB velocity $V$ was defined as the change in the GB separation
$\lambda$ divided by the respective increase in the accumulated
flip time $t_{f}$ between two checkpoints $t_{f}^{\prime}$ and $t_{f}^{\prime\prime}$
\citep{hussein2024model}, 
\begin{equation}
V=-\frac{\lambda(t_{f}^{\prime\prime})-\lambda(t_{f}^{\prime})}{t_{f}^{\prime\prime}-t_{f}^{\prime}}.\label{V_gb}
\end{equation}
Typically, the checkpoint files were saved every $10^{5}$ KMC steps.
The negative sign in Eq.(\ref{V_gb}) accounts for the decrease in
$\lambda$ with time. The reported velocities correspond to time-averaged
values, ensuring that the measured response reflects sustained GB
motion under a constant driving force.

\subsubsection{Solute segregation and solute clustering in grain boundaries}

Solute segregation to the GBs was quantified using a normalized segregation
measure, $\Gamma=c_{\mathrm{gb}}-c_g$, defined as the
difference between the average solute concentrations within the GB
region ($c_{gb}$) and the interiors of the grains ($c_{g}$). The GB region
and grain interiors are illustrated in Figure~\ref{fig:qual_metrics}(b).

To quantify the organization of the solute in the GB regions beyond the mean
 level of segregation $\Gamma$, we analyzed the connectivity of the solute
atoms within the GB region. For each simulation snapshot, we first
constructed the solute distribution map in the GB region using the
indicator $s_{k}=\xi_{k}m_{k}$, where $m_{k}=1$ cells in the GB region and $m_{k}=0$ otherwise. Solute clusters were defined as connected components within the binary
map $\left\{ s_{k}\right\} $ of solute atom positions using nearest-neighbor
connectivity on the underlying lattice. We applied a connected-component
labeling algorithm in which two solute sites belong to the same cluster
if they are connected by a path of solute-occupied sites through neighboring
lattice sites on the square grid. Each connected component
was treated as an individual solute cluster, and its size $S_{\mathrm{C}}$
was measured as the number of solute-occupied lattice sites in that
component. To suppress isolated atoms and transient thermal noise,
clusters smaller than a minimum size threshold ($S_{\mathrm{C}}=4$
lattice sites) were discarded.

For each configuration, we recorded the number of clusters $N_{\mathrm{C}}$
and the distribution of cluster sizes $S_{\mathrm{C}}$. These metrics
enable quantitative characterization of solute aggregation, coarsening,
and cluster breakup in the GB region during driven migration.

\subsubsection{Time averaging and statistical characterization}

For each simulation, the GB velocity, segregation, and clustering metrics
were calculated for a fixed number of snapshots taken from the final
portion of the trajectory and then averaged. We focus on this late-time
window to exclude initial transients and to characterize the long-time
response under the imposed conditions. In this regime, the measured
quantities do not settle to a single constant value at every instant;
instead, they continue to fluctuate in time because GB motion alternates
between periods of reduced mobility (temporary pinning) and renewed
motion (de-pinning), accompanied by ongoing rearrangement of the segregated
solute. The reported averages therefore represent the \emph{mean}
long-time behavior, while the snapshot-to-snapshot variations quantify
the intrinsic dynamical fluctuations of the self-pinning process.

\subsubsection{Quantification of solute drag}\label{subsec:Solute-drag}

The solute drag was quantified by isolating the additional driving
force required to sustain a given GB velocity in the presence of solute
segregation relative to an otherwise identical system in which solute--GB
interactions were disabled. For each thermodynamic condition, two
corresponding sets of simulations were performed: (i) a segregating
system with finite solute--GB interactions, and (ii) a reference
system without solute--GB interactions. In both cases, the GB velocity
$V$ was measured as a function of the applied driving force $F$.

The solute drag force $P$ was defined as the excess force required
to move the GB at a fixed velocity due solely to the effects of the solute. At a given GB velocity $V$, the drag force was calculated as 
\begin{equation}
P(V)=F(V)-F_{0}(V),
\end{equation}
where $F$ and $F_{0}$ denote the driving forces measured in the
segregating and reference systems, respectively.

Because, in general, the simulations do not sample identical GB velocities,
the forces were compared at matched velocities using nearest-neighbor
matching in the velocity space. For each sampled velocity, the closest
velocity in the corresponding reference dataset was identified, and
the difference in the driving force at these matched velocities was taken
as the instantaneous solute drag. To reduce statistical noise and
 obtain a continuous representation of the drag--velocity relationship,
the resulting drag force data were fitted using spline interpolation
as a function of $\log_{10}(V)$ with a regularization term.

The maximum solute drag, $P^{*}$, was identified as the peak of the
smoothed drag curve, 
\begin{equation}
P^{*}=\max_{V}\left[P(V)\right],
\end{equation}
along with the corresponding velocity $V^{\ast}$ at which this maximum
occurred. The values of $F$ and $F_{0}$ reported at $V^{\ast}$
were obtained directly from the raw simulation data at the nearest
sampled velocities to avoid spline-induced bias.

\section{Results}

Before addressing the non-equilibrium processes of solute drag and
GB migration, it is essential to establish the equilibrium thermodynamic
landscape predicted by the model. The structure and stability of solute
atmospheres at a stationary GB set the baseline against which all
kinetic phenomena must be interpreted.

\subsection{Equilibrium segregation and grain boundary phase behavior}

\begin{figure}[ht]
\centering \includegraphics[width=0.55\columnwidth]{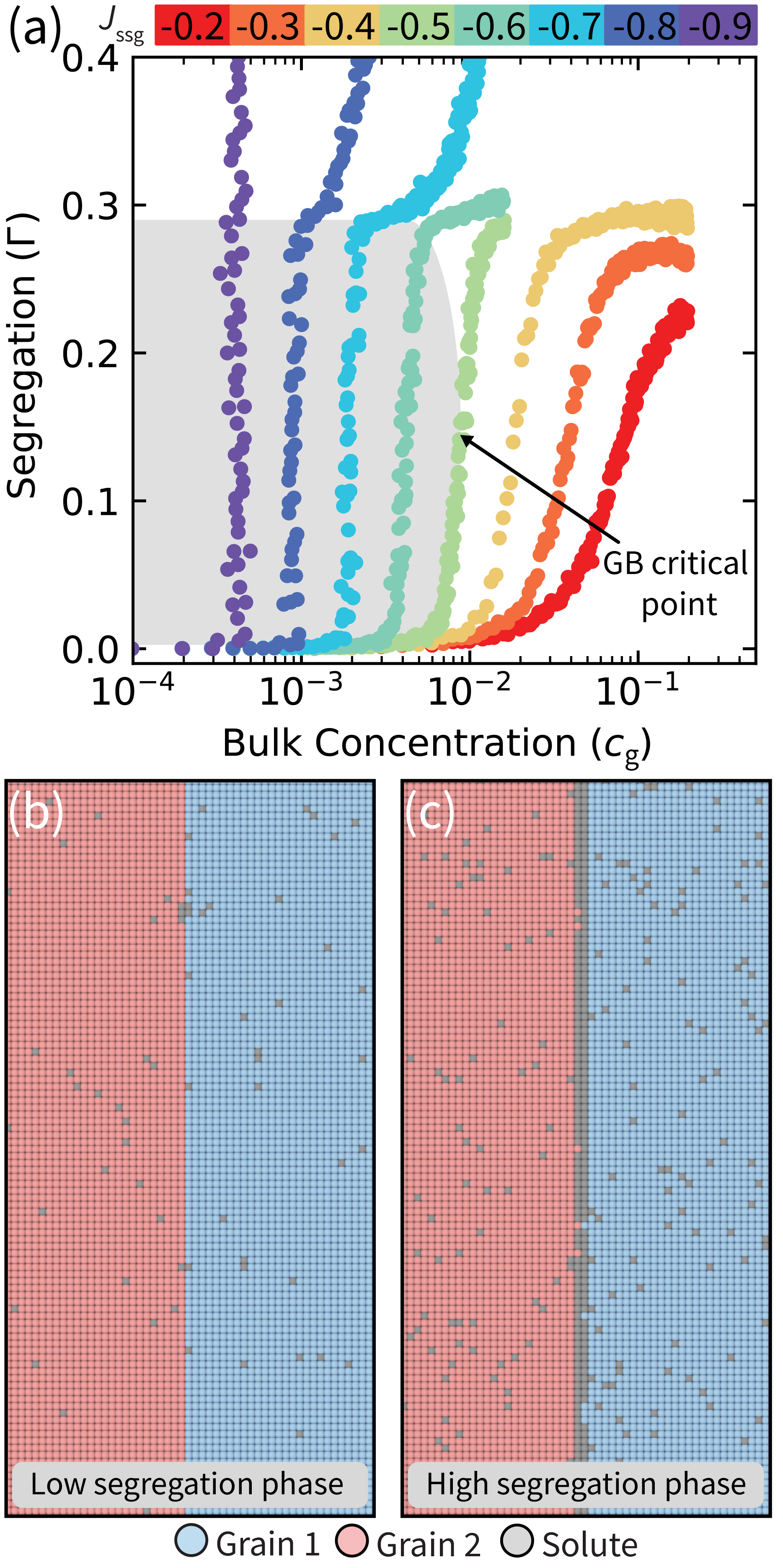} \caption{Equilibrium GB segregation at $T=0.15$ and $J_{\mathrm{sg}}=-0.2$
for selected values of $J_{\mathrm{ssg}}$. (a) Segregation isotherms
showing the amount of GB segregation $\Gamma$ as a function of bulk
solute concentration $c_{g}$. The discontinuities indicate transformations
between two GB phases. (b) Representative snapshot of the low-segregation
GB state at $J_{\mathrm{ssg}}=-0.4$ and $c_{g}=0.01$, characterized
by sparse solute decoration along the GB. (c) Representative snapshot
of the high-segregation GB state at $J_{\mathrm{ssg}}=-0.4$ and $c_{g}=0.05$,
exhibiting a dense, nearly continuous solute-rich segregated layer.}
\label{fig:Tseg_vs_cbulk}
\end{figure}

To understand the thermodynamic origin of the solute clustering, the equilibrium GB segregation was examined as a function of the bulk composition at constant
temperature $T=0.15$ with the attraction of the solute to the GB at $J_{\mathrm{sg}}=-0.2$
and no solute-solute interaction in the grains ($J_{\mathrm{ss}}=0$).
The solute-solute interaction in GBs, $J_{\mathrm{ssg}}$, was 
varied from $-0.2$ to $-0.9$. For each value of $J_{\mathrm{ssg}}$,
the bulk composition $c_{g}$ was varied over an interval in which
the grains remained in a single-phase state, consistent with $J_{\mathrm{ss}}=0$.
The computed solute segregation isotherms $\Gamma(c_{g})$ are shown
in Fig.~\ref{fig:Tseg_vs_cbulk}(a). 

Within the chosen parameter range, phase separation occurs inside
the GB region. At sufficiently strong solute--solute attraction in
the GB, the isotherms exhibit a discontinuity, identifying two distinct
GB phases: a low-segregation phase and a high-segregation phase, as
depicted in Fig.~\ref{fig:Tseg_vs_cbulk}(b,c). As $J_{\mathrm{ssg}}$
becomes less negative, the miscibility gap between the two GB phases
narrows and eventually closes at a GB critical point. The shaded region
in Fig.~\ref{fig:Tseg_vs_cbulk}(a) outlines the miscibility gap
between the two GB phases. Above the critical point, the transition
between the phases becomes continuous, and well above the critical
point the notion of two GB phases loses its significance. 

To study the self-pinning behavior, we chose three values of the solute
interaction parameter $J_{\mathrm{ssg}}$: $-0.2$, $-0.3$, and $-0.4$.
These values are chosen because more negative values of $J_{\mathrm{ssg}}$
lead to a significant widening of the GB and promote the nucleation
of new grains within the system \citep{hussein2024model}. For $J_{\mathrm{ssg}}=-0.3$
and $-0.4$, the segregation isotherms lie slightly above the critical
point, but still display a relatively sharp (albeit not strictly discontinuous)
transition between the low- and high-segregation phases.
This transition allows us to investigate the effect of GB segregation 
on GB dynamics across the full range of segregations from very
low to very high. At low bulk concentrations, the GB adopts a low-segregation
phase characterized by sparse solute decoration (Fig.~\ref{fig:Tseg_vs_cbulk}(b)),
whereas at higher concentrations it transforms into a high-segregation
phase with a dense, nearly continuous solute-rich structure (Fig.~\ref{fig:Tseg_vs_cbulk}(c)).
For $J_{\mathrm{ssg}}=-0.2$, we are significantly above the critical
point and the isotherm describes a gradual transition between the
low- and high-segregation conditions.

\subsection{Grain boundary migration and self-pinning}\label{subsec:GB-migration}

Here we present the results for a bicrystalline alloy with
a bulk solute concentration $c_{g}=0.05$ at a temperature of $T=0.15$.
Grain 1 (center) has a selected orientation $\sigma_{1}$, while
grain 2 has a different orientation $\sigma\neq\sigma_{1}$. The system
is first equilibrated with $J_{\mathrm{sg}}=-0.2$, $J_{\mathrm{ssg}}=-0.4$,
and $J_{\mathrm{s}}=J_{\mathrm{ss}}=F=0$, producing an equilibrium
high-segregation GB phase, as shown in Fig.~\ref{fig:Tseg_vs_cbulk}(c).
The simulation is then performed with finite diffusion kinetics
of the solute by setting the activation barriers to $\varepsilon_{00}^{s}=2$ and
$\varepsilon_{0}^{g}=1$, and applying a synthetic driving force of
$F=0.2$. The GBs are driven toward the center of the simulation cell
as grain 1 shrinks while grain 2 grows, enabling the coupled evolution
of solute segregation and GB motion to be examined.

\begin{figure}[ht]
\centering \includegraphics[width=0.45\columnwidth]{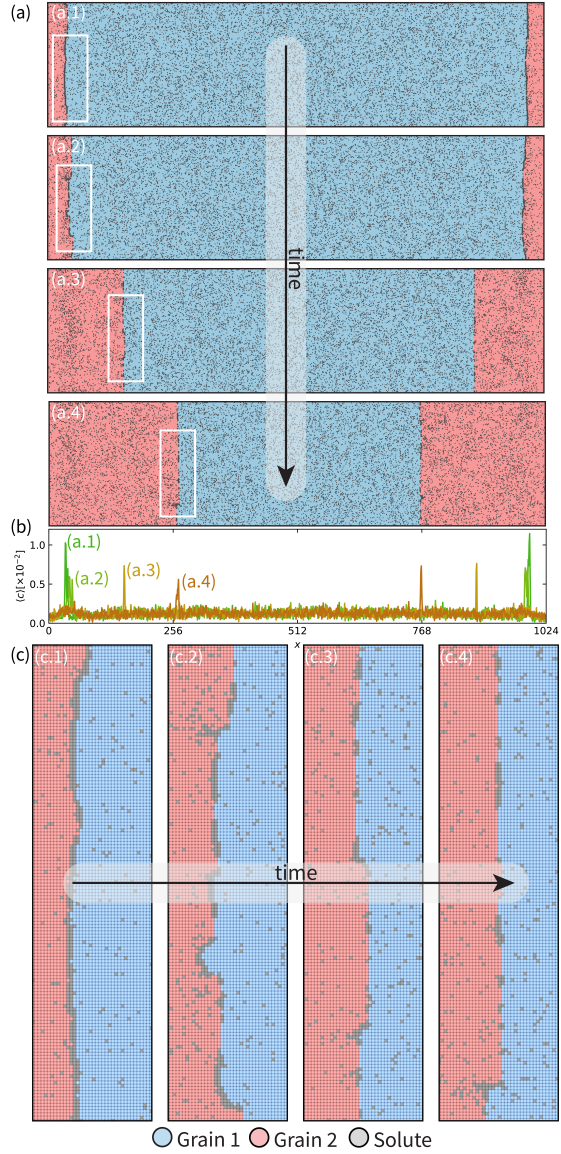} \caption{Self-pinning mechanisms for two initially flat GBs in a binary system
at $T=0.15$, $c_{g}=0.05$, $J_{\mathrm{sg}}=-0.2$, $J_{\mathrm{ssg}}=-0.4$,
$J_{\mathrm{s}}=0$, and $F=0.2$, starting from an equilibrium configuration
with solute segregation. Panels (a.1-4) show the temporal evolution
of the GB configuration: (a.1) Initially, the GB motion is accompanied
by drag of the segregation atmosphere. (a.2) Partial breakup of the
solute atmosphere as the GB advances, accompanied by solute capture,
phase separation, and clustering. (a.3) Pinning of the GB by solute
clusters, leading to the development of local GB curvature and faceting.
(a.4) Evolution into a quasi-steady regime characterized by fluctuations
of the GB velocity and solute cluster morphology about a baseline
configuration. (b) Evolution of the solute concentration profile at
the timepoints of panels (a.1-4). (c) Zoom-in atomistic views of panels
(a.1-4) highlighting the evolution of the GB structure, solute clustering,
and faceting during the self-pinning process.}
\label{fig:self_pinning_qual}
\end{figure}

Figure \ref{fig:self_pinning_qual} illustrates the characteristic
evolution of the two initially flat, segregated GBs under the applied
driving force. At first, the GB migrates while dragging its
segregation atmosphere (Fig.~\ref{fig:self_pinning_qual}(a.1)).
This is evidenced by the high solute concentration peak at the interface
position, which corresponds to the high-segregation GB phase (Fig.~\ref{fig:self_pinning_qual}(b)).
As GB migration continues, the solute atmosphere becomes unstable
and partially detaches from the boundary, leading to local phase separation
and nucleation of solute-rich clusters along the GB (Fig.~\ref{fig:self_pinning_qual}(a.2)).
The larger clusters subsequently pin the boundary, creating a pronounced
GB curvature and faceting (Fig.~\ref{fig:self_pinning_qual} (a.3)).
At longer times, the system evolves into a quasi-steady regime in
which the GB migrates intermittently while interacting with multiple
solute clusters that continuously form, coarsen and dissolve (Fig.~\ref{fig:self_pinning_qual}
(a.4)). This sequence of events demonstrates the emergence of the
self-pinning mechanism in which solute clusters generated \emph{in
situ} at the GB act as effective pinning centers.

\begin{figure}[ht]
\centering \includegraphics[width=0.5\columnwidth]{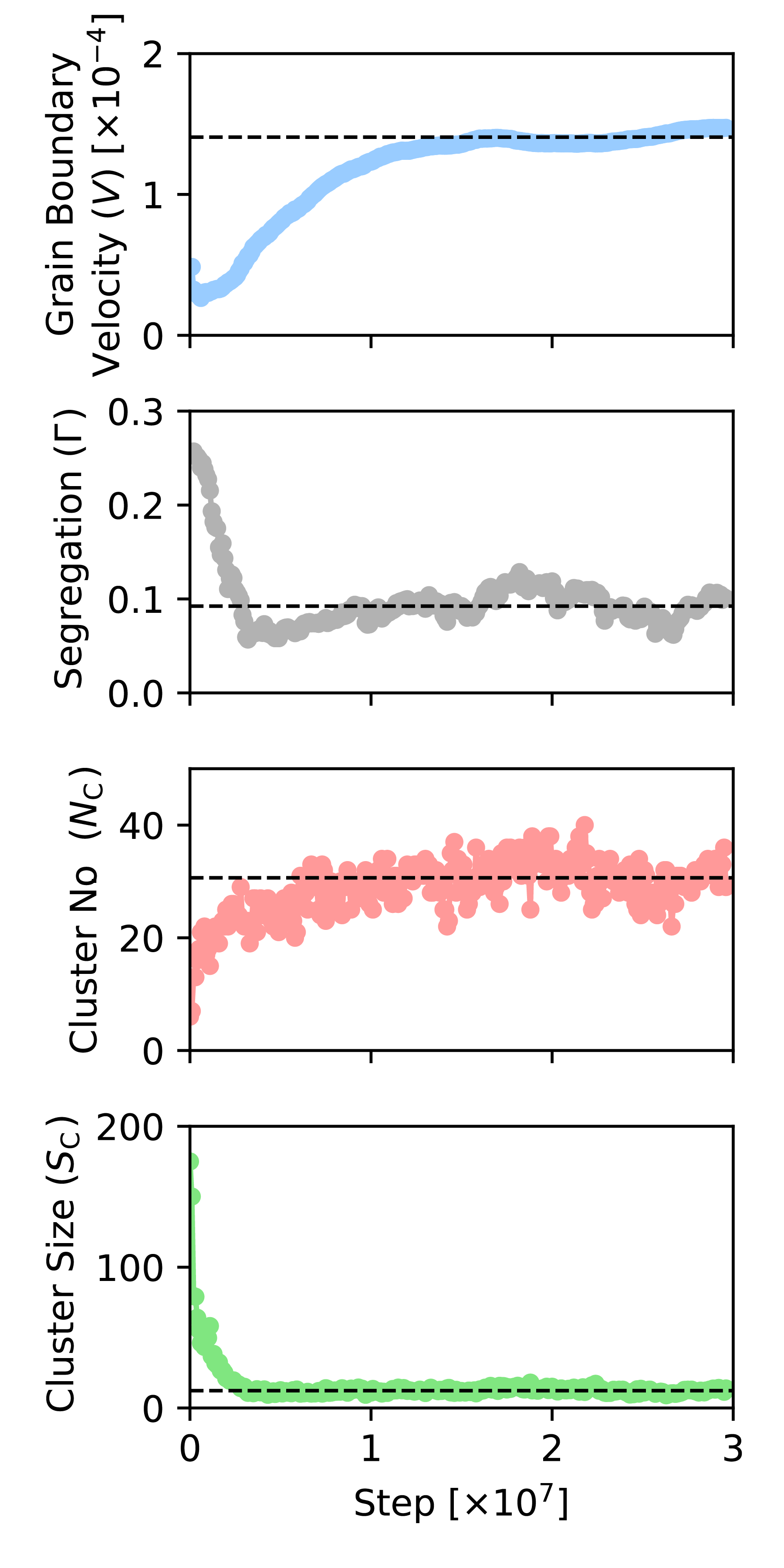} \caption{Quantitative metrics of the system in Fig.\ref{fig:self_pinning_qual}
describing (a) the GB velocity $V$, (b) Segregation $\Gamma$, (c)
Number of clusters ($N_{\mathrm{C}}$) at the GBs, and (d) average
clusters size ($S_{\mathrm{C}}$) at the GBs.}
\label{fig:self_pinning_quant}
\end{figure}

The temporal evolution of quantitative metrics is shown in Fig.~\ref{fig:self_pinning_quant}.
The GB velocity $V$ is initially small due to the classical drag
of the segregation atmosphere but gradually increases under the applied
force as the GB partially separates from the solute-rich environment, as shown
in Fig.~\ref{fig:self_pinning_quant}(a). In the long-time regime,
$V$ fluctuates around a mean value, reflecting intermittent pinning
and de-pinning events. Consequently, the amount of GB segregation $\Gamma$
is initially high due to equilibrium segregation and decreases as
the moving boundary loses part of its solute atmosphere, as shown
in Fig.~\ref{fig:self_pinning_quant}(b). After self-pinning begins,
$\Gamma$ reaches a steady value. The number of GB solute clusters,
$N_{\mathrm{C}}$, increases rapidly due to the breakup of the initially
connected equilibrium cluster and later fluctuates around an average
value; see Fig.~\ref{fig:self_pinning_quant}(c). Meanwhile, the
average cluster size $S_{\mathrm{C}}$ decreases from its initially large
value as the clusters break up; see Fig.~\ref{fig:self_pinning_quant}(d).
This behavior shows that GB self-pinning arises from dynamically evolving
solute-rich clusters that locally hinder GB motion. The dynamic solute
clustering observed during GB migration is directly linked to the
underlying GB phase separation existing on the GB phase diagram. For
weaker solute-solute interactions in the GBs (e.g. $J_{\mathrm{ssg}}=-0.2$),
the segregation isotherm is continuous and clustering is strongly
suppressed.

\subsection{Solute drag and velocity-dependent mobility}

The effect of solute clustering on solute drag, and thus the GB mobility,
was examined by considering several systems with representative interaction
parameters and bulk concentrations. The temperature was fixed at $T=0.15$
and the GB-solute interaction parameter was $J_{\mathrm{sg}}=-0.2$.
The remaining parameter sets were $(J_{\mathrm{ssg}}=-0.2,\,c_{g}=0.150)$,
$(J_{\mathrm{ssg}}=-0.3,\,c_{g}=0.075)$, and $(J_{\mathrm{ssg}}=-0.4,\,c_{g}=0.050)$.

As above, a bicrystalline system was constructed and equilibrated
at zero force following the procedure described in Section~\ref{subsec:GB-migration}.
The initial equilibrium configurations corresponded to points along
the high-segregation branch of the segregation isotherms. Reference
systems with the same bulk solute concentration but without GB--solute
interactions were also constructed to provide the baseline force $F_{0}$.

\begin{figure}[ht]
\centering \includegraphics[width=0.39\columnwidth]{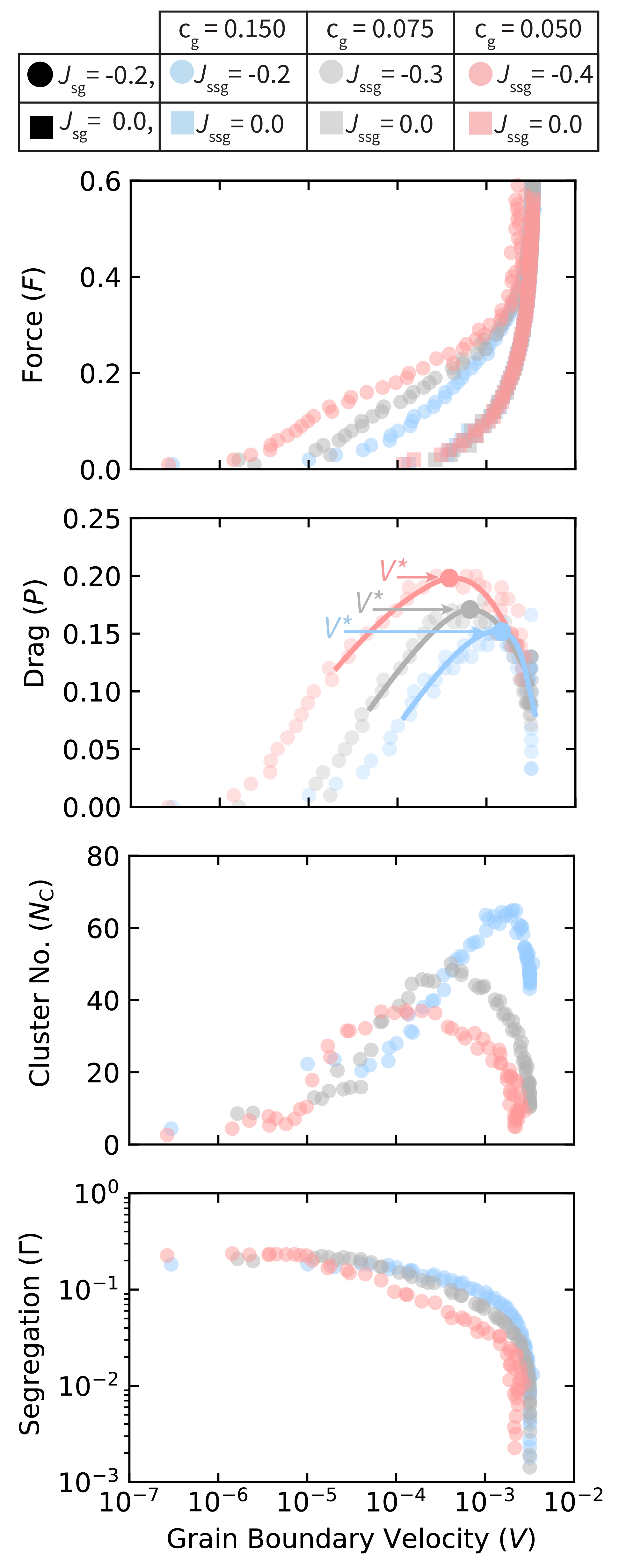} \caption{Coupled evolution of GB mobility, solute drag, solute clustering,
and solute segregation as a function of GB velocity at $T=0.15$ and
$J_{\mathrm{sg}}=-0.2$, shown for three values of the solute--solute
interaction parameter $J_{\mathrm{ssg}}=-0.2$, $-0.3$, and $-0.4$.
(a) Driving force $F$ required to sustain a given GB velocity $V$
for segregating systems (circles) compared to the non-segregating
reference case with $J_{\mathrm{sg}}=J_{\mathrm{ssg}}=0$ (squares).
(b) Solute drag force $P$ exhibiting a pronounced maximum at a characteristic
velocity $V^{\ast}$. (c) Average number of solute clusters along
the GB, $N_{\mathrm{C}}$, showing a peak near $V^{\ast}$ that reflects
the formation and subsequent breakup of solute-induced self-pinning
structures during GB motion. (d) GB segregation, $\Gamma$, as a function
of $V$, demonstrating a monotonic decrease with increasing velocity
and a rapid collapse beyond $V^{\ast}$.}
\label{fig:P_drag}
\end{figure}

Figure~\ref{fig:P_drag}(a) shows the steady-state GB velocity as
a function of the applied driving force $F$. For a given velocity $V$,
segregating systems $(J_{\mathrm{sg}}=-0.2)$ require a larger driving
force than the non-segregating reference system $(J_{\mathrm{sg}}=J_{\mathrm{ssg}}=0)$,
reflecting the solute drag effect. At low driving forces, the difference
between $V(F)$ and $V(F_{0})$ is small. As the force increases,
this difference becomes more pronounced, indicating enhanced drag.
At sufficiently high forces, the GB velocity becomes large enough
to outpace the solute redistribution, suppressing solute drag effects
and leading to converged force--velocity curves.

The solute drag force $P$ was calculated as described in section
\ref{subsec:Solute-drag}. As shown in Fig.~\ref{fig:P_drag}(b),
the function $P(V)$ exhibits a pronounced maximum at a characteristic velocity
$V^{*}$ for $J_{\mathrm{ssg}}=-0.2$, $-0.3$, and $-0.4$. At these
velocities, the GB begins to outrun its solute atmosphere, leading
to the breakup of the uniformly segregated layer into discrete solute
clusters. The average number of clusters $N_{\mathrm{C}}$ peaks near
$V^{*}$ (Fig.~\ref{fig:P_drag}(c)), while the segregation $\Gamma$
decreases monotonically with increasing velocity and drops sharply
beyond $V^{*}$ (Fig.~\ref{fig:P_drag}(d)). At velocities well below
$V^{*}$, solute diffusion maintains pace with slow GB motion, allowing
the boundary to retain a relatively uniform solute atmosphere and
experience only modest drag. At velocities well above $V^{*}$, the
GB effectively detaches from the solute cloud, resulting in reduced
segregation and diminished drag.

The magnitude of the maximum drag force and the location of $V^{*}$
 strongly depend on the solute-solute interaction parameter $J_{\mathrm{ssg}}$.
More negative values of $J_{\mathrm{ssg}}$ enhance the solute clustering,
increase the maximum drag force, and shift $V^{*}$ to lower velocities.
These results demonstrate that GB phase separation and solute clustering
amplify the solute drag and give rise to a robust self-pinning mechanism
that significantly reduces the GB mobility over a finite velocity
range.

\section{Conclusions}

This work identifies self-pinning as an intrinsic stabilization mechanism
for GBs in segregating alloys, arising from GB phase separation and
solute clustering. Using a kinetic Monte Carlo framework that resolves
segregation thermodynamics, diffusion, and GB migration, we have shown
that strong solute--solute attraction at GBs produces first-order
transitions between the solute-lean and solute-rich GB phases. Within
the coexistence regime, a migrating GB spontaneously decomposes into
solute-rich clusters embedded in solute-poor GB regions, generating heterogeneous
GB structures that act as \emph{in situ} pinning centers.

As the GB velocity increases, it progressively outruns its solute
atmosphere, triggering the breakup of uniform segregation into discrete
clusters. This leads to intermittent pinning and de-pinning processes
and produces a velocity-dependent solute drag with a pronounced maximum
at a characteristic velocity $V^{\ast}$. Stronger GB solute--solute
interactions enhance clustering, increase the peak drag force, and
shift the self-pinning regime to lower velocities.

These results demonstrate that the thermodynamic and kinetic stabilization
mechanisms of materials' microstructure are intrinsically coupled:
the same segregation that lowers the GB free energy also generates
solute-rich structures that impede GB motion. Self-pinning, therefore,
provides an effective intrinsic mechanism for suppressing grain growth
without requiring pre-existing second phase inclusions, offering new
routes for designing thermally stable nanocrystalline alloys through
control of GB phase behavior.


\newpage\clearpage{}

\begin{table}[t]
\centering \caption{Physical and normalized variables used in the model.}
\label{tab:variables} %
\begin{tabular}{p{0.55\columnwidth}p{0.12\columnwidth}p{0.18\columnwidth}}
\toprule 
Variable & Physical & Normalized\tabularnewline
\midrule 
Temperature & $T$ & $k_{B}T/J_{gg}$\tabularnewline
Total energy & $E$ & $E/J_{gg}$\tabularnewline
Solute--interface interaction energy & $J_{sg}$ & $J_{sg}/J_{gg}$\tabularnewline
Solute interaction energy in grains & $J_{ss}$ & $J_{ss}/J_{gg}$\tabularnewline
Solute interaction energy in GBs & $J_{ssg}$ & $J_{ssg}/J_{gg}$\tabularnewline
Synthesis force & $F$ & $F/J_{gg}$\tabularnewline
Time & $t$ & $t\nu_{0}$\tabularnewline
\bottomrule
\end{tabular}
\end{table}

\end{document}